\documentclass [manuscript] {acmart}
\usepackage{booktabs} 

\AtBeginDocument{%
  \providecommand\BibTeX{{%
    \normalfont B\kern-0.5em{\scshape i\kern-0.25em b}\kern-0.8em\TeX}}}

\acmDOI{}

\begin{document}

\title{A Study of Comfortability between Interactive AI and Human}

\author{Yi Ru Wang$^*$} 
\affiliation{%
  \institution{University of Washington}
  \city{Seattle}
  \state{Washington}
  \country{USA}
  \postcode{98195}}
\email{yiruwang@cs.washington.edu}

\author{Jiafei Duan$^*$}
\email{duanj1@cs.washington.edu}
\affiliation{%
  \institution{University of Washington}
  \city{Seattle}
  \state{Washington}
  \country{USA}
  \postcode{98195}
}

\author{Sidharth Talia}
\affiliation{%
 \institution{University of Washington}
  \city{Seattle}
  \state{Washington}
  \country{USA}
  \postcode{98195}}

\author{Hao Zhu}
\affiliation{%
  \institution{University of Washington}
  \city{Seattle}
  \state{Washington}
  \country{USA}
  \postcode{98195}}
\email{haozhu@cs.washington.edu}

\def\thefootnote{*}\footnotetext{Equal contribution}


\renewcommand{\shortauthors}{Wang et al}

\begin{abstract}
As the use of interactive AI systems becomes increasingly prevalent in our daily lives, it is crucial to understand how individuals feel when interacting with such systems. In this work, we investigate the comfort level of individuals when interacting with intent-predicting AI systems and identify the factors of influence. We introduce a study protocol to analyze human comfortability when interacting with intent-predicting AI systems and execute the study with over a dozen participants. The study findings suggest that users are comfortable with AI systems if they have control and their privacy is not affected. Additionally, the study found that users could differentiate between AI and human responses, but this did not significantly affect their comfort levels. This research paper's significance lies in its contribution to the growing body of literature on interactive AI systems, and it emphasizes the need to consider user perceptions in the development and deployment. 

\end{abstract}

\maketitle

\section{Introduction}
\begin{figure}
  \includegraphics[width=\textwidth]{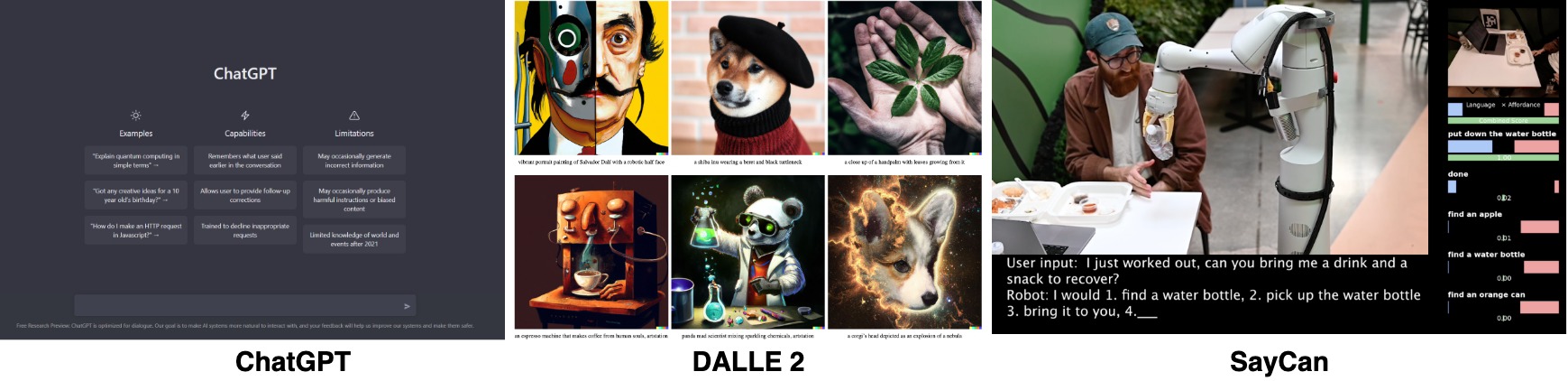}
  \caption{Left: \textbf{ChatGPT}, a large language model trained by OpenAI that uses natural language processing to generate responses to questions and statements. Middle: \textbf{DALLE 2}, an AI system developed by OpenAI that can create realistic images and art from a description in natural language Right: \textbf{SayCan}, a household robot model developed by Google Robotics to perform household tasks via natural language instruction.}
  \label{fig:intro}
\end{figure}

At what point do AI (Artificial Intelligence) systems interacting with humans start being unnerving? Do humans have a preference for AI systems that refrain from making sensitive predictions? How comfortable are humans with AI system's capability to make predictions? With the diversification of predictive capabilities \cite{chowdhery2022palm,ho2020denoising,duan2022survey,ahn2022can} of AI methods (shown in Figure \ref{fig:intro}) as a result of recent advances in deep learning \cite{lecun2015deep}, embodied AI \cite{duan2022survey}, large language models (LLMs) \cite{devlin2018bert,chowdhery2022palm}, and generative models \cite{dhariwal2021diffusion,ramesh2021zero} (examples shown in Figure 1), these questions have become especially relevant. Hence, it is a crucial time for us to evaluate the comfortability of humans interacting with such interactive AI systems. Hence, using of the proposed AI system that is trained to predict human intent through perception, we aim to evaluate the comfortability of the user when interacting with such a system. We hypothesize that such an attempt may cause discomfort \cite{zhong2022augmenting} to the human. To evaluate this hypothesis, we perform a user study with an AI system equipped with Theory of Mind (ToM) \cite{premack1978does} that is capable of predicting human intention in a social context described in \cite{duan2022boss}. 

We aim to use the \cite{duan2022boss} AI system to conduct a comparison study in which the participant performs a designated set of social interactions -- described in \cite{duan2022boss} -- with an interface equipped with a screen and camera. On the other side of the screen, either the AI system or a human observer will be making a prediction of the participant's intent. After the prediction, the participant will then be asked a series of questions regarding the interaction. For the purpose of this study, we have recruited 13 participants from various departments across the university to take part in our user study. More specifically, we will explore and attempt to address the following research questions from our user study:
\begin{itemize}
    \item \textbf{How} comfortable are humans in interacting with interactive AI systems that possess the capability to infer their intentions. 
    \item \textbf{What} are the factors that affect the comfortability of humans when interacting with such an interactive AI system.
    \item \textbf{What} information can we learn from the user study that could be beneficial to improve the development of more human-centric and interactive AI systems that people are comfortable in interacting with.
\end{itemize}

To address these questions, we conducted a user study with 13 participants and interviewed them before and after the study. Our contributions include: (a) designing and conducting the experimental protocol for the user study to determine whether humans are comfortable with an AI system that can predict their intentions. (b) conducting pre- and post-interviews with the subjects and gathering pertinent data for further analysis. (c) analyzing and drawing new insights from the data gathered, and providing recommendations for enhancing the comfort of future Human-AI interactions.

\section{Related Work}

This work builds on top of prior work that explores the Theory of Mind, Modelling of human thoughts and work that looks into user concerns about the ever growing encroachment of AI in their day-to-day lives.

\noindent\textbf{Theory of Mind (ToM).} Premack et. al \cite{premack_woodruff_1978} state that the Theory of mind is the ability to attribute beliefs, desires, capabilities, goals, and mental states to others. This allows humans to reason about others’ thoughts and behaviors, and anticipate them. Studies \cite{woolley, engel} have shown that the ToM capabilities of a team’s members are a strong predictor of team performance. Çelikok et. al \cite{ccelikok2019interactive} argue that the results of ToM from human-human and agent-agent interaction also apply to human-AI interaction. Our work investigates the comfortability of humans when interacting with AI equipped with a theory of mind. 

\noindent \textbf{Modeling human thoughts.} The works \cite{doshi2010modeling, baker2011bayesian, de2013much, han2018learning}
proposed various methods such as using empirically informed POMDP (Partially Observable Markov Decision Process), bayesian theory of mind, agent-based simulation, etc. for modeling human thoughts. The works \cite{baker2008theory, baker2014modeling, sukthankar2014plan, kleiman2016coordinate, rabinowitz2018machine} propose theory-based methods of different kinds to tackle social goal inference. In recent years, learning-based methods have also gained traction. The works \cite{wunder2011using, hadfield2016cooperative} have proposed modeling human thoughts as a multi-agent problem either with iterated reasoning or inverse reinforcement learning (IRL). The works \cite{bobu2021feature_missing, bobu2022inducing} further study how IRL can unearth the value functions of humans through corrective feedback. The work in \cite{duan2022boss} proposes a benchmark for human belief prediction given context of the objects. In this work, we use the dataset from \cite{duan2022boss} to train a CNN to predict the human's beliefs.

\noindent \textbf{User Concerns.} The works \cite{Kalimeri2020,Maria} study the concerns that people may have regarding using AI for daily activities. In general, they observe that users' opinions depend on their dependency on AI, and the associated interference and surveillance that such systems involve. \cite{HAII} mentions that while users enjoy the convenience brought on by AI, they are hesitant to have such systems make decisions for them, or for the systems to have a "mind of its own". Further, users are concerned about the collection of their data by such systems. This is reflected by the shift of users toward systems that do not hold user data by default \cite{Snapchat}. While this does not preclude the use of AI -- as models could still be trained with temporary data -- it demonstrates that general users don't want to be "explicitly" tracked. In this work,  we investigate the concerns that users have regarding interactive AI which can predict their intentions through a user-study protocol.


\begin{figure}
  \includegraphics[width=\textwidth]{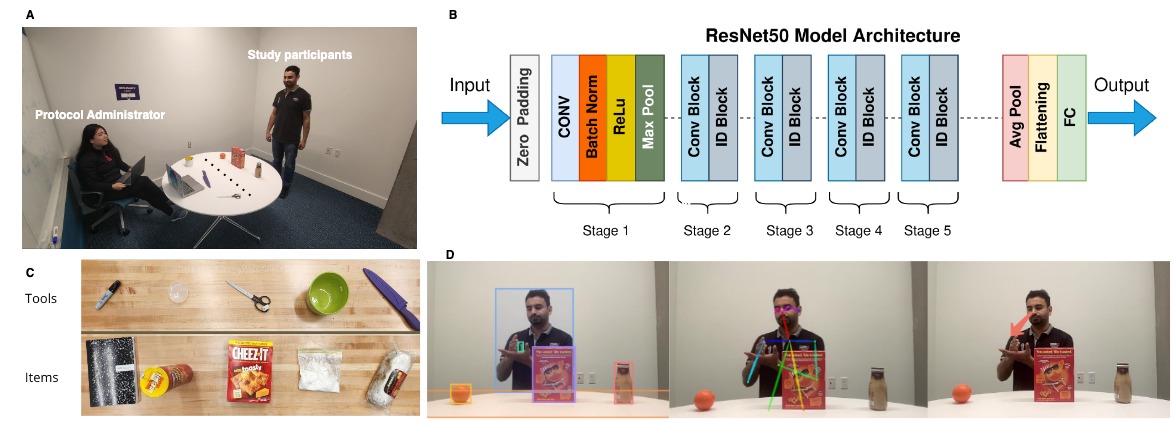}
  \caption{The overview of the study protocol. (A) Environment setting of the user study. (B) Model Architecture of ResNet50 which was used for training. (C) The finalised list of object-tool pairs. (D) The different training modalities input into the pretraining model.}
  \label{fig:fig2}
\end{figure}

\section{Methods \& Design Protocol}
\label{sec:methods}
For the design of the user study, we mimic the same setting as mentioned in \cite{duan2022boss}, modified to exclude the interaction between the second participant and the first participant. Each participant is directed to sit in front of a table, on top of which there are a series of 5 contextual object-tool pairs. We set up a laptop connected via zoom to a remote location as depicted in Figure \ref{fig:fig2}. The contextual objects and tools are $o_{i}^{object}$= \{\emph{Chips}, \emph{Book}, \emph{Bread}, \emph{Crackers}, \emph{Sugar}\} and, $o_{i}^{tool}$=\{ \emph{Knife}, \emph{Bowl}, \emph{Scissors}, \emph{ChipsCap}, \emph{Marker}\} respectively. The idea of the study is for the individual to describe the tools using non-verbal gestures, the video of which will be captured and transferred to the AI system or person on the other side of the screen. For example, if the participants selects the "marker", they would likely gesture the action of drawing over their palms, which mimics the action of writing words in a book. The demonstration of the participants is streamed either to an AI system or a human, whose prediction response will be revealed to the participant through a text that appears on the laptop screen. The participants repeat this process 4 times with a different object each time. Upon completion of the study, the participants will then be asked a series of post-interview questions. The AI system used here for intent prediction is a CNN with a ResNet50\cite{he2016deep} backbone trained on the dataset collected from \cite{duan2022boss}.
\begin{figure}
  \includegraphics[width=\textwidth]{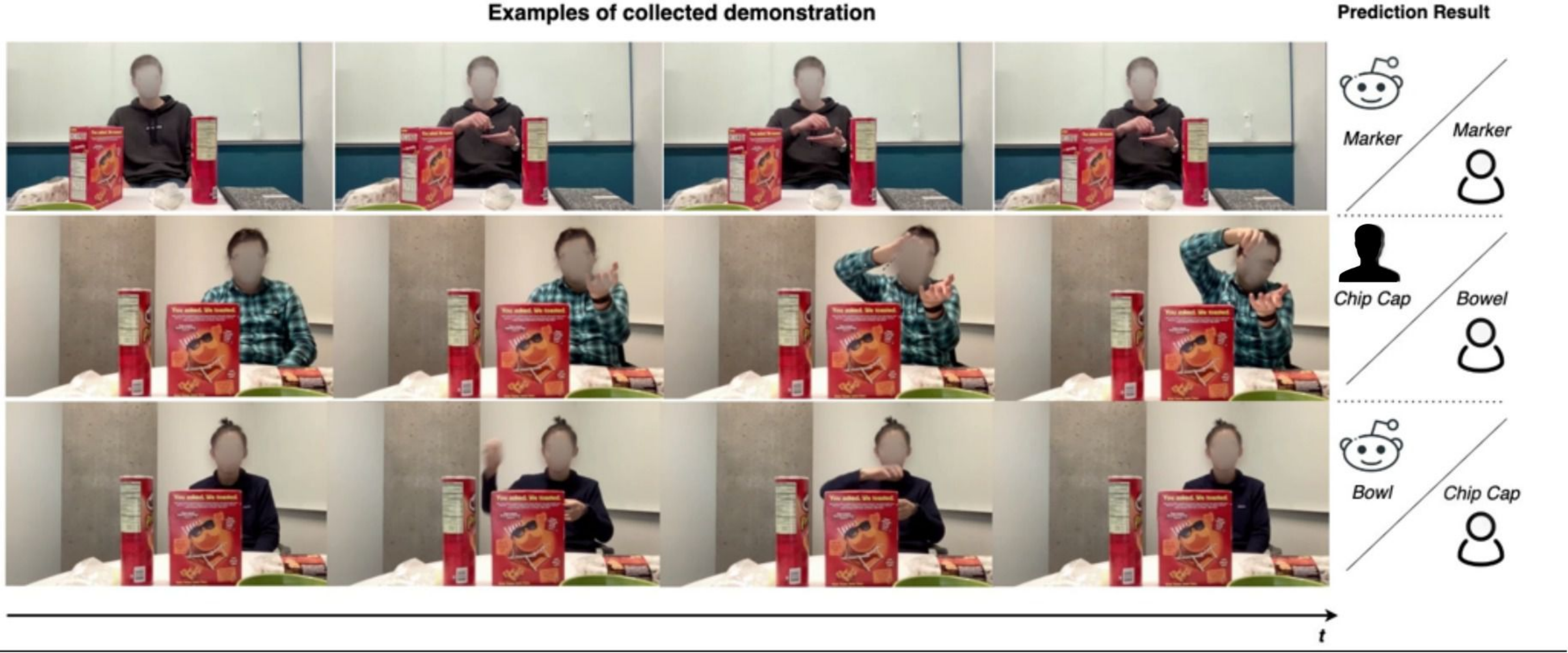}
  \caption{Examples of the collected demonstration from our user study and the respective prediction result by either AI or human.}
  \label{fig:teaser}
\end{figure}

\section{Study Results \& Analysis}

\subsection{User Study Setup}
Participants were recruited on-site from a single academic institution, and were composed of graduate/undergraduate students from different fields as well as managerial staff. Strict adherence to the approved IRB protocol was enforced, and hence all participants were made aware of the video collection component of the experiment. 
For executing the study, each participant was scheduled for a private 30-minute session, which begun with a brief introduction of the experimental procedure, including setup, objects involved, expectations of the participant, and the data that will be collected. During this time, the participants were encouraged to ask any clarification questions. To ensure familiarity with the experimental flow, all participants engaged in a practice run prior to participating in the real experiment. The real user study consisted of five objects and tool pairs, as highlighted in Section \ref{sec:methods}. Hence, we task the participants to gesture for a total of four tool-object pairs, to avoid the trivial fifth tool-object pair, following the procedure described in Section \ref{sec:methods}. During the experiment, we ask the participants to rate the response speed of each answer on a scale of 0 (much slower than human) to 4 (much faster than human). Note that we take precaution to not mention to the participant the possibility of interaction with an AI system to prevent bias. Upon completion of the experiment, the participants are tasked with answering a few questions related to their prior experience in interacting with AI systems, engagement throughout the experiment, and general thoughts about interacting with an AI system equipped with a ToM. A selection of questions is shown in Table \ref{tab:survey-questions}.



\begin{table}[htbp]
  \centering
  \caption{Survey questions}
  \begin{tabular}{@{}p{0.05\textwidth}p{0.6\textwidth}p{0.25\textwidth}@{}}
    \toprule
    \textbf{\#} & \textbf{Question} & \textbf{Response format} \\
    \midrule
    \multicolumn{3}{c}{Background Information} \\
    \midrule
    Q1 & How often do you interact with AI systems on a daily basis? & Scale (1-5) \\
    Q2 & What is the last time you interacted with an AI system and for what purpose? & Open-ended \\
    Q3 & Have you ever felt uncomfortable interacting with an AI system? Why? & Open-ended \\
    Q4 & What are some key attributes which you believe would make AI systems more comfortable to interact with? & Open-ended \\

    \midrule
    \multicolumn{3}{c}{Post-Experiment Questions} \\
    \midrule

    Q5 & How comfortable were you throughout the experimental process? & Scale (0-5) \\
    Q6 & Which responses do you think were from the AI system? & Multiple choice \\
    Q7 & How concerned are you about AI that can predict your thoughts? & Scale (0-5) \\
    
    \bottomrule
  \end{tabular}

  \label{tab:survey-questions}
\end{table}

\subsection{Analysis of Participants' Backgrounds}
\begin{figure}[h]
  \includegraphics[width=\textwidth]{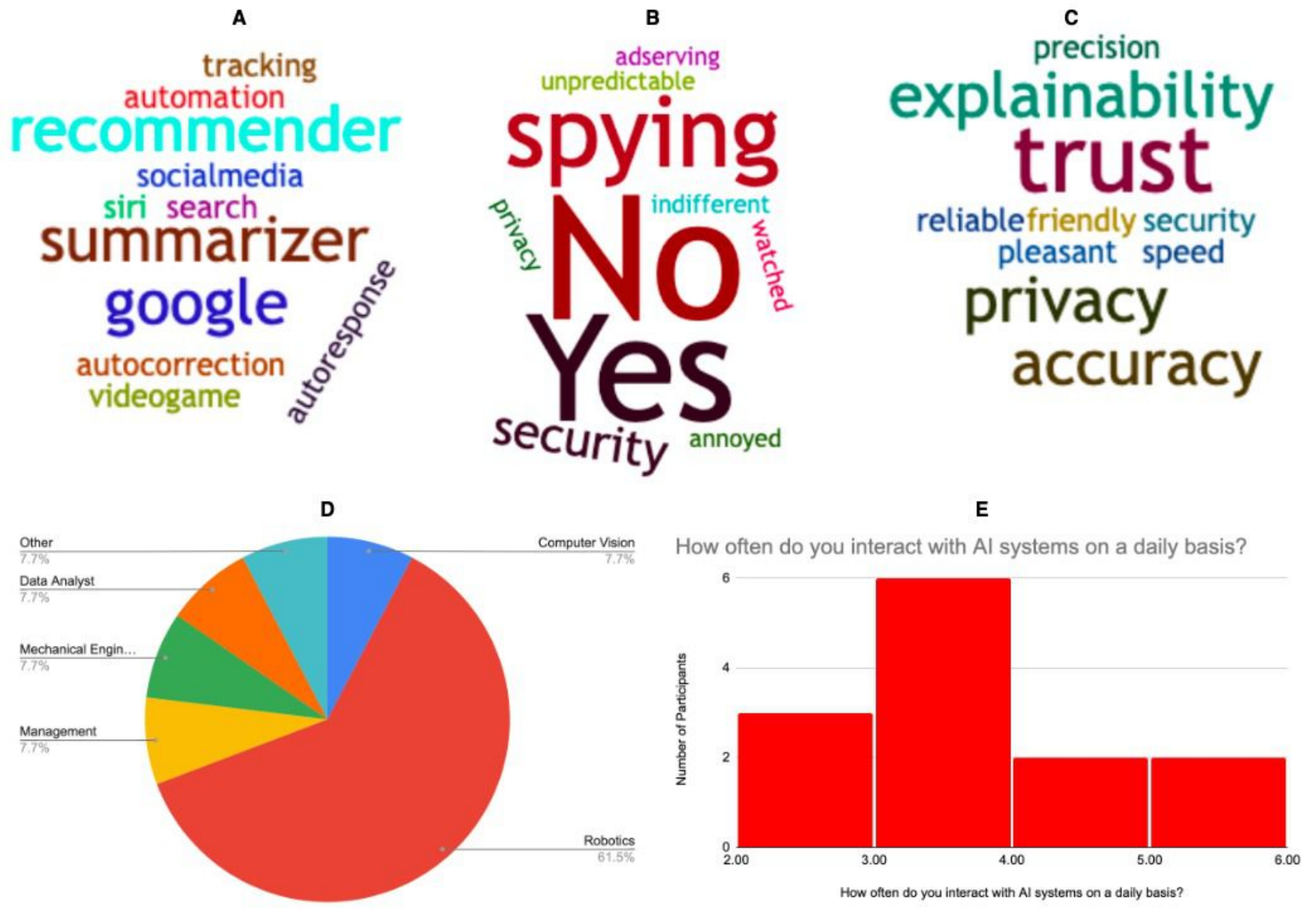}
  \caption{\textbf{Pre-study survey results 1.} A) Response to most recent interaction with AI system; B) Response to incidents where participants are most concerned about interacting with an AI system; C) Factors which participants would believe would make AI systems more comfortable to interact with; D) General background information about the participants; E) A polling result on how often the participants think they interact with an AI system on a daily basis.}
  \label{fig:pre-study}
\end{figure}

We conduct an analysis of the participants' backgrounds from several dimensions: current occupation, frequency of interaction with AI, primary modes of interaction, main concerns in interacting with AI, and features in which they wish to see in AI systems. While looking at the background of participants, as shown in Figure \ref{fig:pre-study}D, we see that the majority of our participants have a robotics background (61.5\%), and (69\%) of participants' work involve AI on a regular basis (Robotics and Computer Vision). This means that the results of the study may be most applicable to drawing conclusions for the population that heavily interacts or are familiar with AI systems. Examining the frequency of interaction with AI systems, as shown in Figure \ref{fig:pre-study}E, we find that all participants interact with AI systems more than twice a day. The majority of participants interact with AI for approximately thrice a day. Out of all participants, the average number of interactions is 3.25, with a standard deviation of 1.01. Note that the plot refers to conscious interaction, and repetitive or continuous interaction with the same AI system is considered as a single interaction. In terms of the most recent interaction with an AI system which participants can recall, as shown in Figure \ref{fig:pre-study}A, we see that most indicated interactions with recommendation systems, search systems, or voice assistants. Participants were primarily concerned with spying / tracking of their activities, and the security / privacy of their data, as seen from Figure \ref{fig:pre-study}B. One of the participants responded to the question \textit{"Have you ever felt uncomfortable interacting with an AI system? Why?"} with \textit{"Yes. It gives me the feeling i am being watched all the time"}. When asked which factors would make AI more comfortable to interact with, most participants mentioned improvements in privacy, trust, and reliability, as shown in Figure \ref{fig:pre-study}C. They hope to see systems where the user's data is kept secure, where the data is not tracked without permission, and where the system can perform intended functions with accuracy.

\subsection{Analysis of Study Results}

\begin{figure}
  \includegraphics[width=\textwidth]{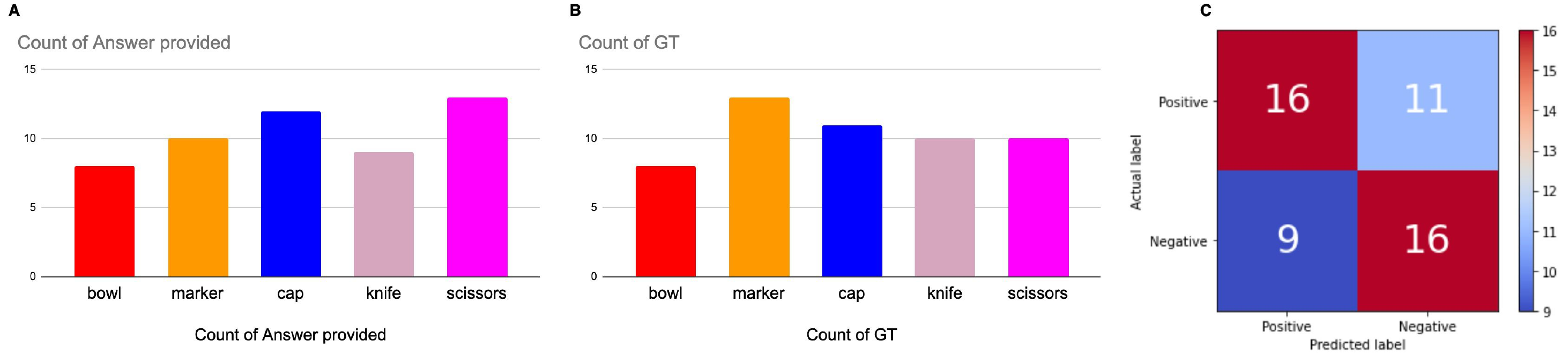}
  \caption{\textbf{Study Analysis} A) Count of answer provided for each tool. B) Count of GT predicted for each tool. C) A breakdown of the result matrix for the prediction.}
  \label{fig:data-distribution}
\end{figure}

\textbf{Distribution of Results.} In examining the distribution of outputs from the model in comparison to the ground truth objects chosen, as can be seen from Figure \ref{fig:data-distribution}, the ground truth actions performed is relatively even, with approximately ten demonstrations per object. However, it can be seen that more participants preferred to demonstrate marker in comparison to demonstrating bowl. The AI system's output is distributed fairly evenly, with a count hovering around ten for most tools. 

\noindent \textbf{Performance of AI System vs. Human Observer.} Comparing the performance of the AI system with that of the human observer, we find that humans in general have a higher prediction precision (89\%) than AI systems (64\%), as seen in Table \ref{table:three_sections}. We also find that AI predictions and human observer predictions differ in prediction time, as perceived by the user. As seen in Table \ref{table:three_sections}, we see that on average, the speed of AI predictions was on average 0.8, while the speed of human predictions was 1.8. Note that the criteria for ranking the speed is that 0 indicates slow response, while 2 indicates the average response speed of a human. 

\noindent \textbf{Differentiation between AI System vs. Human Observer.} In examining the ability of participants to differentiate between an AI system response and a human response, we find that participants are able to predict correctly 61.5\% of the time, as shown in Table \ref{table:three_sections}. From qualitative experimental observations, the basis of differentiation from AI and humans used by the majority of participants is the speed of prediction and the prediction accuracy. Most participants associated slow response and incorrect predictions with an AI system, since they believed that humans should respond quickly and with near-perfect accuracy. We further analyze the participant predictions of the backend system with a confusion matrix, as shown in Figure \ref{fig:data-distribution}C. We find that for 30.8\% of predictions, the users are able to distinguish that a prediction is generated by AI. As well, for 30.8\% of predictions, the users are able to distinguish that a prediction is made by a human. In 17.3\% of the cases, the users mistakenly classified an AI prediction as a human prediction, and in 21.3\% of the cases, they interpreted a human prediction as an AI prediction. This shows that in general, users are capable of differentiating a prediction made by an AI system from a human prediction. Only in rare cases, do they confuse AI predictions with human predictions.


\begin{table}[ht]
\centering
\begin{tabular}{|c|c|c|c|c|c|} \hline
\multicolumn{2}{|c|}{Precision of prediction} & \multicolumn{2}{|c|}{Speed of prediction} & \multicolumn{2}{|c|}{Users' prediction of the backend} \\ \hline
Subject & Success Rate (\%) & Subject & Speed of prediction (0-4) & Correctness & Success Rate (\%)  \\ \hline
AI & 64.0 & AI & 0.8 & Correct & 61.5 \\
Human & 88.9 & Human & 1.8 & False & 38.5\\ \hline
\end{tabular}
\caption{Result Analysis: Precision of prediction of AI responses vs human responses. Speed distribution of AI responses vs human responses. Users' prediction of the backend.}
\label{table:three_sections}
\end{table}

\begin{figure}
  \includegraphics[width=\textwidth]{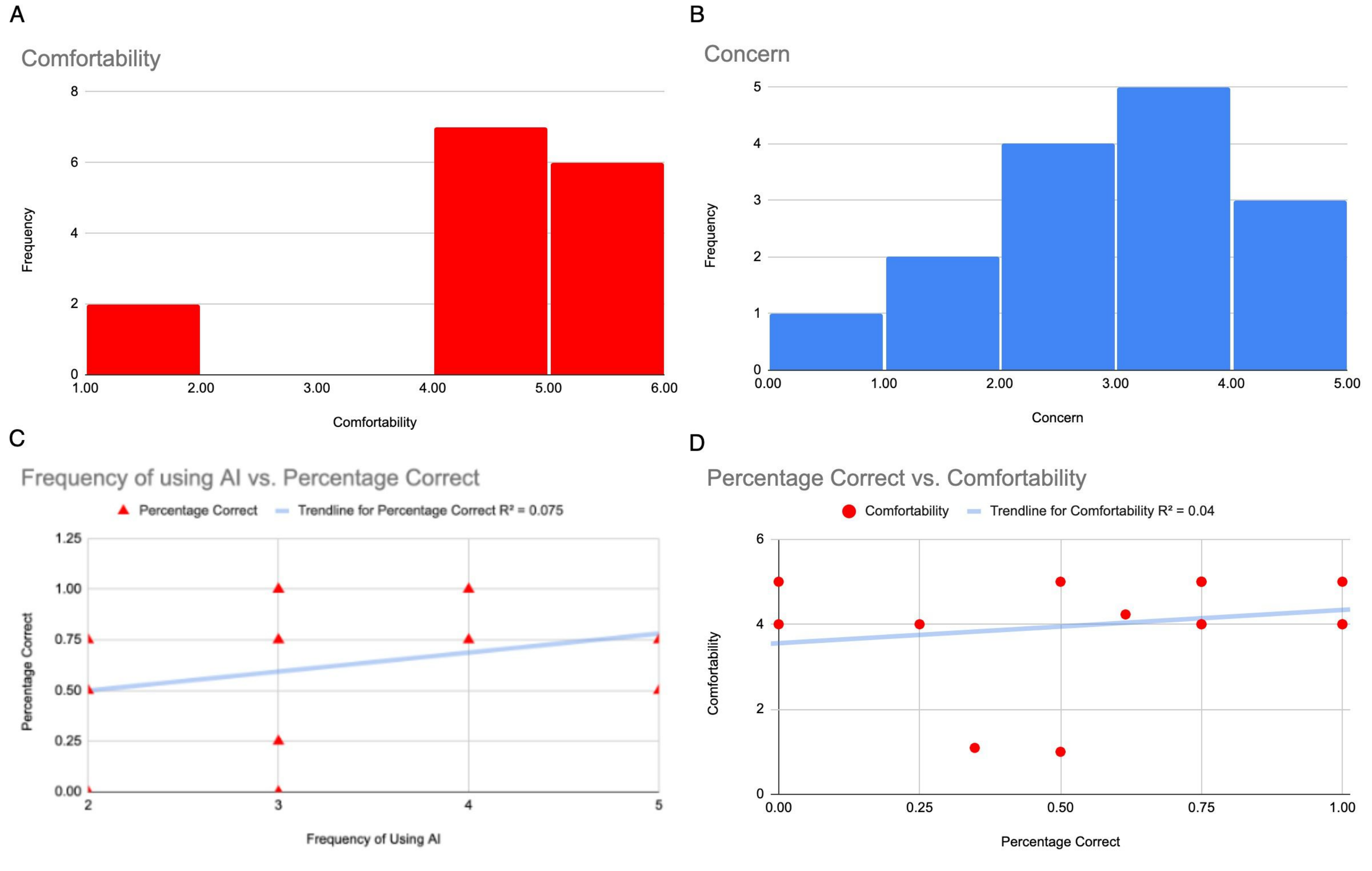}
  \caption{\textbf{Post-study Analysis} A) Frequency plot on how comfortability the participants are when dealing with our AI system. B) Frequency plot on how concern the participants are after interacting with our AI system. C) Plot on percentage correct against comfortability. D) Plot on percentage correct against concern.}
  \label{fig:comfort-concern}
\end{figure}

\noindent \textbf{Comfortability and Concern.} In examining the overall comfortability and concern that participants had for AI systems after participating in the user study, we find that the majority of participants are comfortable with engaging with our AI system, and not overly concerned about AI systems with a ToM. As shown in Figure \ref{fig:comfort-concern}A, the majority of the users indicated high comfortability throughout the experiment ($4.2 \pm 1.1$). Only a single user reported low comfortability when interacting with the AI system, which through clarifications, was due to a misunderstanding of the rating scale. From Figure \ref{fig:comfort-concern}B, we find that many participants indicated minor concerns towards general AI with a ToM ($2.6 \pm 1.2$). In interpreting the results, it must be noted that the sample size consists of primarily people working in computer-science-related fields. The fact that most participants interact with AI systems on a regular basis could be a contributing factor to the overall observation that people are comfortable and have minor concerns when interacting with AI. We also conducted an analysis to examine whether the ability of participants to differentiate between AI and human predictions correctly correlated with their comfortability and concern towards AI. As shown in Figure \ref{fig:comfort-concern}C and \ref{fig:comfort-concern}D, we see that there is a low coefficient of determination for both Comfortability vs. Percentage Correct, and Concern vs. Percentage Correct. This means that the variance in the dependent variables (comfortability and concern) is not influenced by the independent variable (percentage of correct predictions of backend system).

\section{Conclusion}
In this work, we began our study with the question of how comfortable humans are with having interactive AI systems in their day-to-day lives which can predict their intent and what factors affect their comfort when interacting with such systems. To this end, we designed a study protocol and conducted a user study with 13 participants from within the University of Washington, Seattle. Overall, results revealed that participants were relatively comfortable with AI systems that could infer their intentions, as long as their data was kept private and under control. Participants of the study excelled at predicting whether the predictor was a human or an AI, using speed and correctness as primary signals for differentiation. However, it must be noted that because the differentiation was being made using speed and correctness, and a system with better speed and accuracy may have yielded a different result. We would recommend that when designing future interactive AI systems, AI practitioners should focus on aligning the model inference speed and accuracy with that of humans to enable a natural interaction. Potential future extensions of the work include expanding and diversifying participant demographics, which will yield more generalizable results and conclusions which are applicable to the general population. Another potential future extension involves development and refinement of the AI model to improve accuracy, prediction speed, and diversity of prediction scope to handle a wider range of behaviour. This could involve incorporating additional data sources and using better models to learn interaction from the data. Overall, we hope that the study protocol and results presented in the work can facilitate future research in understanding the comfortability of humans in interacting with AI systems with a theory of mind and enable AI systems that are comfortable to interact with.

\section*{Acknowledgements}

The authors thank James Fogarty and Lisa Elkin for fruitful discussions and feedback. We would also like to acknowledge the University of Washington Institutional Review Board (IRB) for granting exemption \#STUDY00016679 for the study.

\bibliographystyle{ACM-Reference-Format}
\bibliography{sample-base}










\end{document}